%% file: 1mainbody.tex
\documentclass[aip,amsmath,amssymb,reprint,floatfix]{revtex4-2}

\usepackage{graphicx}
\usepackage{dcolumn}
\usepackage{bm}
\usepackage[utf8]{inputenc}
\usepackage[T1]{fontenc}
\usepackage{mathptmx}
\usepackage{etoolbox}

\input{packages_commands}

\draft 

\begin{document}

\title{Brillouin-based storage of QPSK signals with fully tunable phase retrieval} 

\author{Olivia Saffer}
\thanks{Authors contributed equally to this work.}
\author{Jes\'us Humberto \surname{Marines Cabello}}
\thanks{Authors contributed equally to this work.}
\author{Steven Becker}
\author{Andreas Geilen}
\author{Birgit Stiller}
\email[Author to whom correspondence should be addressed: ]{\mbox{birgit.stiller@mpl.mpg.de}}
\homepage[Group homepage: ]{\url{optoacoustics.de}}
\affiliation{Max Planck Institute for the Science of Light, Staudtstr. 2, 91058, Erlangen, Germany}
\affiliation{Department of Physics, Friedrich-Alexander-Universit\"at Erlangen-N\"urnberg, Staudtstr. 7, 91058, Erlangen, Germany}

\date{30 September 2024}

\begin{abstract}
Photonic memory is an important building block to delay, route and buffer optical information, for instance in optical interconnects or for recurrent optical signal processing. Photonic-phononic memory  based on stimulated Brillouin-Mandelstam scattering (SBS) has been demonstrated as a coherent optical storage approach with broad bandwidth, frequency selectivity and intrinsic nonreciprocity. Here, we experimentally demonstrated the storage of quadrature-phase encoded data at room temperature and at cryogenic temperatures.  We store and retrieve the 2-bit states $\QPSKbit$ encoded as optical pulses with the phases $\QPSK$ --- a quadrature phase shift keying (QPSK) signal. The 2-bit signals are retrieved from the acoustic domain with a global phase rotation of $\pi$, which is inherent in the process due to SBS. We also demonstrate full phase control over the retrieved data based on two different handles: by detuning slightly from the SBS resonance, or by changing the storage time in the memory scheme we can cover the full range $[0,2\pi)$. At a cryogenic temperature of 3.9~K, we have increased readout efficiency as well as gained access to longer storage times, which results in a detectable signal at 140~ns. All in all, the work sets the cornerstone for optoacoustic memory schemes with phase-encoded data.
\end{abstract}

\maketitle

\input{introduction}

\input{concepts}
\input{setup}

\input{measurements_results}
\input{conclusion}

\section*{\MakeUppercase{Supplementary material}}
The supplementary material includes the following: measures against setup drift; details of the data analysis; further plots of the storage time sweep in \cref{subsec:lifetime}; and a discussion on the anomaly at $\tstore = 6$ ns for the detuned memory measurements in \cref{subsec:tunedcryoQPSK}.

\section*{\MakeUppercase{acknowledgments}}
We acknowledge funding from the Max Planck Society through the Independent Max Planck Research Groups scheme, the DFG project STI 792/1-1 and the Max Planck School of Photonics. 

The authors thank Kevin Jaksch, Stefan Richter, Laura Bl\'azquez Mart\'inez and Grigorii Slinkov for very helpful discussions and experimental support.

\section*{\MakeUppercase{Author declarations}}
\subsection*{Conflict of interest}
The authors have no conflicts to disclose.

\subsection*{Author contributions}
\noindent \textbf{Olivia Saffer}: Data curation (equal); Investigation (equal); Methodology (equal); Validation (equal); Writing - original draft (lead); Writing - review and editing (equal). \textbf{\mbox{Jes\'us Humberto Marines Cabello}}: Data curation (equal); Formal analysis; Investigation (equal); Methodology (equal); Software (lead); Validation (equal); Visualisation; Writing - review and editing (equal). \textbf{Steven Becker}: Conceptualization (supporting); Data curation (supporting); Methodology (supporting); Software (supporting); Writing - review and editing (equal). \textbf{Andreas Geilen}: Conceptualization (supporting); Data curation (supporting); Methodology (supporting); Software (supporting); Writing - review and editing (equal). \textbf{Birgit Stiller}: Conceptualization (lead); Funding acquisition; Project Administration; Resources; Supervision; Writing - original draft (supporting); Writing - review and editing (equal).

\section*{\MakeUppercase{Data Availability}}
The data that support the findings of this study are available from the corresponding author upon reasonable request.

\clearpage

\newpage

\appendix
\onecolumngrid

\maketitle This Supplementary Material provides additional analyses and
discussions on the following topics: (\ref{ssec:drift}) countermeasures for setup drift; (\ref{ssec:dataanalysis}) details of the data analysis and further plots of the storage time sweep data in \cref{subsec:lifetime}; and (\ref{ssec:readout_anomaly}) a discussion of the anomaly in frequency-detuned measurements in \cref{subsec:tunedcryoQPSK} at a storage time of 6~ns.

\raggedbottom
\input{drift}
\input{dataanalysis}

\input{readoutanomalystudy}
\nocite{} 
\newpage

\hrule

\end{document}

%% file: packages_commands.tex
\usepackage{xfrac, nicefrac}

\usepackage{gensymb}
\usepackage{titlesec}

\newcommand{\pr}{{\,\prime}}
\newcommand{\lb}{\left}
\newcommand{\rb}{\right}
\DeclareMathSymbol{\shortminus}{\mathbin}{AMSa}{"39}

\newcommand{\omegaD}{\omega_\text{\,D}}
\newcommand{\omegaC}{\omega_\text{\,C}}
\newcommand{\DOmega}{\Delta\Omega}

\newcommand{\tstore}{t_\text{\,store}}

\newcommand{\phiD}{\phi_\text{\,D}}
\newcommand{\phiR}{\phi_\text{\,R}}

\newcommand{\QPSK}{\{0,\frac{\pi}{2},\pi,\frac{3\pi}{2}\}}
\newcommand{\QPSKbit}{\{00,01,10,11\}}

\newcommand{\AuCI}{\text{\,AuC}_I}
\newcommand{\AuCQ}{\text{\,AuC}_Q}

\usepackage{float}

\usepackage{tikz}

\usepackage{cleveref}
\crefname{figure}{Fig.}{Figs.}
\Crefname{figure}{Figure}{Figures}
\crefname{table}{Table}{Tables}
\crefname{equation}{Eq.}{Eqs.}
\crefname{section}{Sec.}{Secs.}
\crefname{page}{p.}{pp.}
\crefname{appendix}{Appendix}{Appendixes}

%% file: introduction.tex
\section{\MakeUppercase{Introduction}}
\label{sec:intro}
Optical storage and buffer techniques are crucial for advancing all-optical networks and computing systems, which provide higher processing speed, broader bandwidth and lower energy consumption than their electronic counterparts. These techniques can reduce the latency in optical interconnects by overcoming the von Neumann bottleneck and are an essential requirement when it comes to neuromorphic computing. Optical information can be encoded in amplitude and phase using different frequency channels, polarization modes or even orbital angular momentum modes and a photonic memory ideally stores at least a subset of these parameters. Types of optical memory include fiber delay lines,\cite{Brunner2018,Vandoorne2014,Xu2021} slow-light based techniques,\cite{Tucker2005,Mirzaiee2022,Gong2023} optical microresonators,\cite{Hill2004,Fiore2013,DelBino2021} semiconductor optical amplifier (SOA)-based schemes\cite{Pleros2009,Pitris2016,Tsakyridis2019} and many more.\cite{Alexoudi2020,Lian2022,Kari2023} Recently, the domain of acoustics was used for delaying and storing optical information based on stimulated Brillouin-Mandelstam scattering (SBS).\cite{Merklein2018,Eggleton2019} There are different techniques based on SBS, such as Brillouin slow light,\cite{Okawachi2005,Song2005,Yi2007,Pant2012} quasi light storage,\cite{PreuSSler2009,Jamshidi2012,Merklein2024} Brillouin dynamic gratings,\cite{Chin2012,Song2009} and photonic-phononic memory,\cite{Zhu2007,Kalosha2008,Merklein2017} also called simply Brillouin-based memory. The latter transfers the information from the fast optical domain to slow traveling acoustic waves and back via pulsed SBS interactions. This concept is coherent,\cite{Merklein2017} frequency preserving,\cite{Stiller2019} intrinsically nonreciprocal,\cite{Dong2015,Merklein2021} can be operated as a cascade\cite{Stiller2018} and down to 100~ps short pulses.\cite{Stiller2023,Piotrowski2021} The limitation in terms of storage time based on the acoustic lifetime was tackled by actively reinforcing the acoustic waves \cite{Stiller2020} or operating the system in a cryogenic environment.\cite{Geilen2023} The coherence of the Brillouin-based memory is of particular interest for enhancing the optical information density by using amplitude and phase encoding and is therefore predestined to store and retrieve the phase space of information. To date, the coherence of Brillouin-based memory has not yet been used to its full potential. As pointed out in a recent theoretical study of Brillouin-based memory,\cite{Nieves2021} previous works have focused mainly on the intensity of the readout rather than the phase. While the phase of the readout has been investigated using single homodyne detection,\cite{Merklein2017,Stiller2019} a full experimental investigation of the phase using double balanced homodyne detection was done for the first time only very recently.\cite{Geilen2023} 

In this work, we experimentally demonstrate the storage and retrieval of a quadrature phase-shift keying (QPSK) signal as a proof-of-concept. We store and retrieve the 2-bit signals $\QPSKbit$ encoded as optical pulses with the phases $\QPSK$, both at at room temperature and at cryogenic temperatures of around 4~K. By carefully calibrating the optical phase, we show that the retrieved data acquires a global phase rotation of $\pi$. We also demonstrate and experimentally investigate in detail the additional feature of Brillouin-based memory --- exceptional phase control of the read out data.  The phase of the retrieved data can be all-optically manipulated from 0 to 2$\pi$ in two ways in the memory scheme: (i) by detuning from the SBS resonance at a fixed storage time or (ii) by changing the storage time at a fixed frequency detuning. Both methods are implemented by changing the frequency and timing of the control pulses, which leaves the data unaltered and places full control onto the counterpropagating write and read pulses. At a cryogenic temperature of 3.9~K, we are able to detect the retrieved data up to 140~ns after the original data pulse. 

Brillouin-based memory is not limited to storing four distinct phase states, as in QPSK, but could be enhanced for higher order encoding such as 16 Quadrature Amplitude Modulation (16QAM) signals or further. The demonstrated complete control over the phase of the retrieved data can prove very useful in advanced optical signal processing schemes and neuromorphic network implementations.\cite{Becker2024,Bacvanski2024} This enhances the body of work surrounding Brillouin-based memory and paves the way for optical memory schemes for phase-encoded data.

%% file: concepts.tex
\section{\MakeUppercase{Concept}}
\label{sec:briefconcepts}

SBS\cite{Brillouin1922,mandelstam1926light,Wolff2021} is a third order nonlinear optical process that couples two optical waves at frequencies $\omega_\text{\,P}$ (input pump wave) and $\omega_\text{\,S}$ (scattered wave) and one acoustic wave with the frequency $\Omega$. The frequency of the scattered waves can either be downshifted ($\omega_\text{\,S} = \omega_\text{\,P} - \Omega$) or upshifted ($\omega_\text{\,S} = \omega_\text{\,P} + \Omega$). The frequency difference $\Omega$ is referred to as the Brillouin frequency shift (BFS). The scattering process conserves energy and momentum and is coherent. \cite{Boyd2003,Agrawal2019,Wolff2021} 

\begin{figure}
    \centering
    \includegraphics{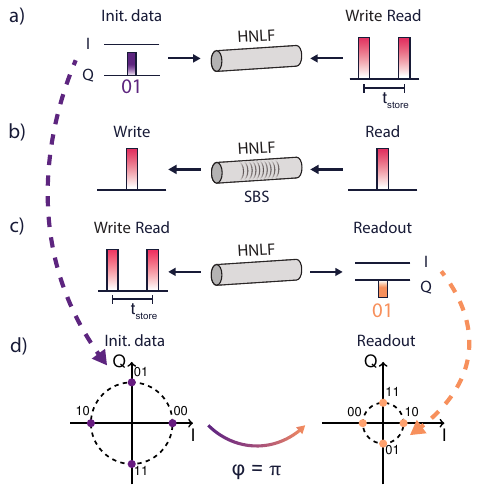}
    \caption{The principle behind coherent optical information storage. (a) We encode the two-bit states $\QPSKbit$ as data pulses with four different phases $\QPSK$, respectively. The 01 state is shown as an example. This is sent into the sample, a section of highly nonlinear fiber (HNLF), from one side while the write control pulse gets sent in from the other side. (b) Stimulated Brillouin-Mandelstam scattering (SBS) creates a coherent acoustic wave in the HNLF sample when the data and write pulses meet inside the fiber. The write pulse and depleted data pulse exit the HNLF from opposite sides before the read pulse enters the sample after a time $\tstore$. This time delay of $\tstore$ defines the storage time of the initial data. (c) The read pulse interacts with the written acoustic wave and creates readout pulses. (d) The readout pulses gain a global phase rotation $\varphi = \pi$ with respect to the phases of the initial data, but the phase differences between the four signals are preserved. The arrows illustrate how the 01 data and readout pulses correspond to the constellation maps.}
    \label{fig:conceptsQPSKmemory}
\end{figure}

Brillouin-based memory\cite{Zhu2007,Kalosha2008,Dong2015,Merklein2017,Stiller2018,Stiller2019,Stiller2020,Merklein2021,Geilen2023,Stiller2023,Piotrowski2021} allows the transfer of optical information from the optical to the acoustic domain. In the memory scheme, two counterpropagating optical pulses are involved: data and control. The data is at frequency $\omegaD$ and the control at frequency $\omegaC = \omegaD - \Omega$. The memory consists of a write step and a read step. In the write step, the write control pulse interacts with the counter-propagating data pulse (\cref{fig:conceptsQPSKmemory}(a)). This depletes the data pulse and creates a coherent acoustic wave via SBS (\cref{fig:conceptsQPSKmemory}(b)). In the read step, the read control pulse interacts with the previously written acoustic wave and transfers the information back into an optical pulse, the readout pulse, at the same frequency as the original data pulse (\cref{fig:conceptsQPSKmemory}(c)). As depicted in \cref{fig:conceptsQPSKmemory}(a), the time between the respective midpoints of the write and read control pulses determines the storage time, $\tstore$. 

Since the SBS process is coherent, there is also a fixed phase relation between the data and readout. In a previous work\cite{Geilen2023} it was found that the phase of the data and readout is conserved, with the readout gaining a phase shift $\varphi~=~\pi$, i.e., if the data has phase $\phiD$ then the readout will have the phase $\phiR =\phiD + \pi$. 

This makes Brillouin-based memory suitable for storing data where the phase is important, such as when phase-key digital modulation techniques are used to encode data. QPSK, and its variants, is a widely used digital modulation technique to encode transmitted data as 00, 01, 10 or 11, using both in-phase (I) and quadrature (Q) components of the signal. The phases of these four two-bit states are separated by multiples of $\frac{\pi}{2}$. 

In our scheme for storing coherent optical information, we encode the QPSK states $\QPSKbit$ in data pulses with the phases $\QPSK$, respectively. The arrow from \cref{fig:conceptsQPSKmemory}(a) to \cref{fig:conceptsQPSKmemory}(d) illustrates this mapping of phase to two-bit state for the 01 state and the arrow from \cref{fig:conceptsQPSKmemory}(c) to \cref{fig:conceptsQPSKmemory}(d) shows the subsequent readout. \Cref{fig:conceptsQPSKmemory}(d) summarizes the change from initial data to the retrieved data with a global rotation $\varphi~=~\pi$.

%% file: setup.tex
\section{\MakeUppercase{Experimental setup}}
\label{sec:setup}

\begin{figure*}[t!]
    \centering
     \includegraphics{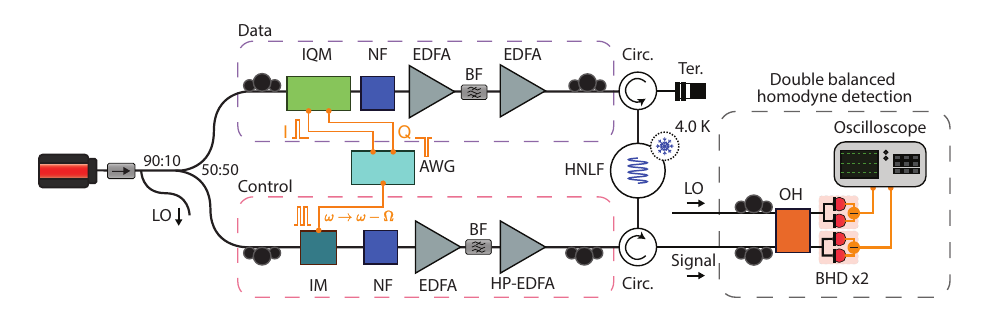}
    \caption{Experimental setup. On the control arm, the laser is downshifted by $\Omega$ and the write and read pulses are generated. On the data arm, the calibration and data pulses are generated. In the diagram the configuration is shown for the 10 signal: the calibration pulse is sent on $I$ and the 10 data pulse on $Q$. The two arms of the setup go via circulators into one of two samples of highly nonlinear fiber (HNLF): one sample is at room temperature and the other in a cryostat at around 4~K. Measurements are performed using double balanced homodyne detection. LO: local oscillator; IQM: intensity-quadrature (IQ) modulator; IM: intensity modulator; NF: narrow bandpass filter; EDFA: erbium-doped fiber amplifier; AWG: arbitrary waveform generator; BF: broad bandpass filter; HP-EDFA: high power erbium-doped fiber amplifier; Circ.: circulator; HNLF: highly nonlinear fiber; Ter.: beam terminator; OH: $90\degree$ optical hybrid; BHD: balanced homodyne detector.}
    \label{fig:setupdiagram}
\end{figure*}

A simplified diagram of the experiment is given in \cref{fig:setupdiagram}.  A narrow-linewidth 1550~nm laser is split into two arms: data and control. Before this split, there is an additional coupler to create the local oscillator (LO) arm at frequency $\omega_\text{\, laser} = \omegaD$. 

The light in the control arm is sent into an intensity modulator (IM) where an arbitrary waveform generator (AWG) is used to simultaneously shift the frequency to $\omegaC = \omegaD - \Omega$ and create rectangular 2~ns write and read pulses, separated by a time $\tstore$. A narrow bandpass filter ensures that only light at $\omegaC$ is transmitted. The control pulses are amplified using erbium-doped fiber amplifiers (EDFAs) to an average power of approximately 1~W, before being sent into the sample.  

On the data arm, an intensity-quadrature modulator (IQ modulator) is used together with the AWG to create rectangular 2~ns pulses with a chosen phase $\phiD$. The data arm is also amplified using EDFAs to an average power of approximately 1\% of the average control power before being sent into the sample. An additional rectangular 2~ns calibration pulse with phase zero precedes the data pulse. This pulse is timed so that it travels through the sample before the control pulses arrive and does not interact with the read and write control pulses. Hence, the calibration pulse is not involved in the memory process. The purpose of this pulse is to find the arbitrary phase between signal and LO. 

The repetition rate of the pulses on each arm is 5 MHz. The pulse waveforms are synchronised so that the control write and read pulses interact with data pulse and acoustic wave within the central part of the sample. 

Since SBS is polarization sensitive, the polarization of the data and control pulses have to match. For this reason, the control and data arms have polarization controllers before the light is sent into the sample. We used two samples of germanium-doped highly nonlinear fiber (HNLF) - one at room temperature and another in a cryostat at either 4.0 K or 3.9 K. The sample at room temperature is 2~m long. The sample in the cryostat has a total length of 2.7~m, of which 2.5~m are at the cryogenic temperature. 

Before detection, a narrow bandpass filter is used to filter out any reflections of the control pulses. Detection is done via double balanced homodyne detection, which consists of a $90\degree$ optical hybrid, two 2.5~GHz balanced detectors and a 4~GHz oscilloscope. Two additional polarization controllers are used for the signal and LO before going into the hybrid, to maximize the signal on the oscilloscope. 

The two channels of the oscilloscope that correspond to the two outputs of the balanced homodyne detectors we denote $I$ and $Q$, and the oscilloscope traces captured from these channels we denote $I(t)$ and $Q(t)$, respectively.

%% file: measurements_results.tex
\section{\MakeUppercase{Measurements and results}}
\label{sec:meas&results}

\subsection{Storage and retrieval of QPSK signals at room temperature}
\label{subsec:roomtempQPSK}
\input{roomtempQPSK}

\begin{figure}[t!]
    \centering\includegraphics{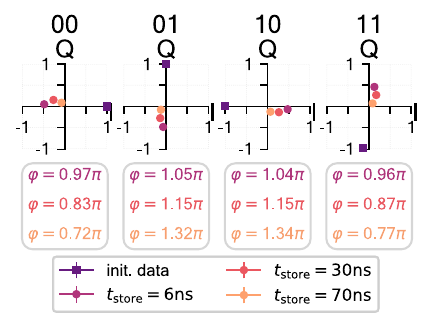}
    \caption{Analogously with \cref{fig:RT_QPSK_IQ}(a), the initial four QPSK data signals $\QPSKbit$ with the phases $\QPSK$ are compared with their respective readouts for the three storage times. These measurements were performed at a temperature of 4.0~K. }
    \label{fig:4Kstorage}
\end{figure}

\begin{figure*}[t!]
    \centering
    \includegraphics{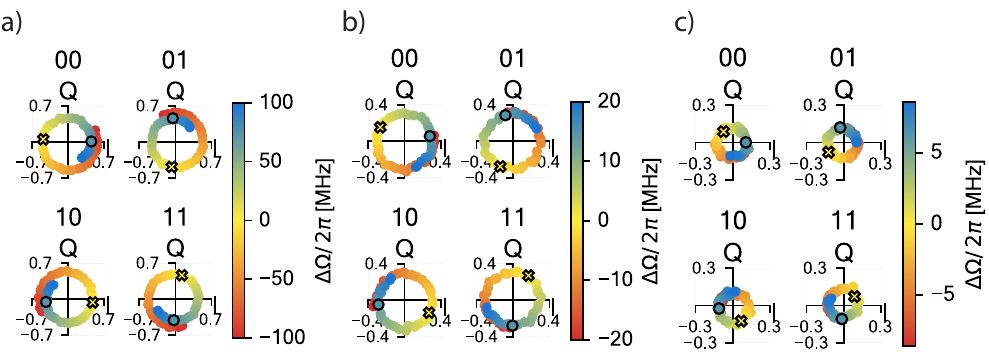}
    \caption{By detuning the frequency of the control pulses by $\Delta\Omega$, we can control the readout phase, $\phiR$. In the IQ plots above, the outlined cross represents the frequency matched case where $\Delta\Omega = 0$. The outlined circle represents the frequency detuning for which $\phiR~\approx~\phiD$. Note that for longer storage times, the frequency range of $\Delta\Omega$ required to cover the full phase space is decreased. (a) Storage time of 6~ns and frequency detuning range of $\DOmega/2\pi = \pm~100.0$~MHz. The outlined circle corresponds to $\Delta\Omega/2\pi$~=~75.0~MHz. The non-perfect overlap at the extremes of the frequency detuning is due to a modulator artifact present in the oscilloscope traces. This is discussed in detail in the supplementary materials, \cref{ssec:readout_anomaly}. (b) Storage time of 30~ns and frequency detuning range of $\DOmega/2\pi = \pm~20.0$~MHz. The outlined circle corresponds to $\Delta\Omega/2\pi$~=~13.0~MHz. (c) Storage time of 70~ns and frequency detuning range of $\DOmega/2\pi = \pm~8.6$~MHz. The outlined circle corresponds to $\Delta\Omega/2\pi$~=~4.0~MHz. Note that the axes for (a)-(c) have been scaled to match the readout AuCs.}
    \label{fig:detunedIQ}
\end{figure*}

\begin{figure*}[t!]
    \centering
    \includegraphics{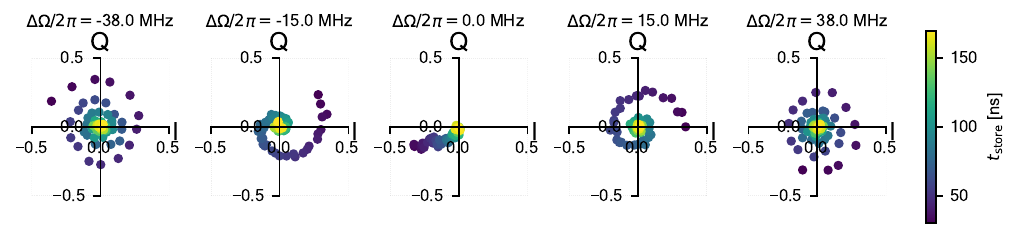}
    \caption{In these plots, it can be seen how detuning the frequency of the control pulses affects the readout phase, $\phiR$, as we sweep over the storage time, $\tstore$. These measurements were performed for the 00 state at 3.9~K. In \cref{fig:detunedIQ}, the storage time was fixed and the control frequency detuning, $\Delta\Omega$, was swept. Here we choose a particular $\Delta\Omega$ and sweep $\tstore$ from 30~ns to 170~ns. For the frequency matched case ($\Delta\Omega$~=~0), the readout decays to zero without $\phiR$ undergoing much variation. However, for positive detunings $\phiR$ spirals anti-clockwise as the readout decays towards zero, and for negative detunings $\phiR$ spirals clockwise as the readout decays towards zero. }
    \label{fig:detunedlifetimeIQ}
\end{figure*}

\subsection{Storage and retrieval of QPSK signals at 4.0~K}
\label{subsec:cryoQPSK}
\input{cryoQPSK}

\subsection{Phase tuning of retrieved data pulses}
\label{subsec:tunedcryoQPSK}
\input{tunedcryoQPSK}

\subsection{Storage time sweep at 3.9~K}
\label{subsec:lifetime}
\input{lifetimecryomemory}

%% file: roomtempQPSK.tex
\begin{figure}[b]
    \centering
    \includegraphics{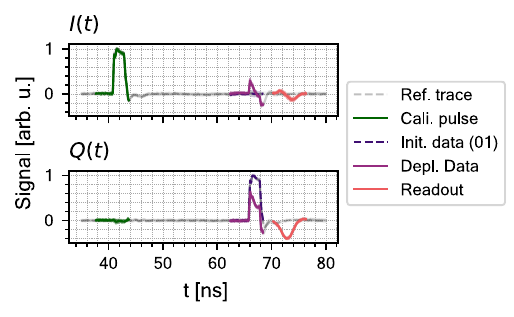}
    \caption{Example mean $I(t)$ and $Q(t)$ traces measured on the oscilloscope for \mbox{$\tstore = 6$ ns} and the 01 state. The mean reference trace is plotted with a dashed line. The calibration pulse with phase $\phi~=~0$, in green, precedes the 01 data pulse. The initial data pulse, in dark purple, can be compared with the depleted data pulse in light purple. Since this pulse corresponds to $\phiD~=~\frac{\pi}{2}$, it is almost completely in the $Q(t)$ trace and only slightly in the $I(t)$ trace. This is also true for the corresponding readout, in orange, which has the phase \mbox{$\phiR \approx \phiD + \pi$}.}
    \label{fig:traces}
\end{figure}

\begin{figure}
    \centering
    \includegraphics{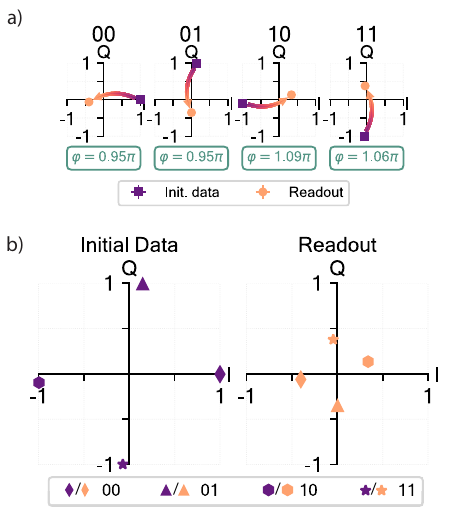}
    \caption{The initial four QPSK data signals $\QPSKbit$ with the phases $\QPSK$ are compared with their respective readouts for a storage time of $\tstore = 6$~ns, at room temperature. The coordinates on the IQ diagrams are given by $\lb(\AuCI, \AuCQ\rb)$, where \mbox{$\text{\,AuC}_{I}$ $\lb(\text{\,AuC}_{Q}\rb)$} is the area under the curve calculated from the \mbox{$I(t)$ $\lb(Q(t)\rb)$} oscilloscope traces. (a) Here it can be seen that the readout phases have a shift $\varphi \approx \pi$ with respect to the data phases i.e.  $\phiR \approx \phiD + \pi$. (b) The initial data signals are plotted on a single set of axes to show that the phases of the different signals differ by multiples of $\frac{\pi}{2}$. This is also true for the retrieved signals (the readouts). The signals gain a global $\pi$ rotation, but this does not affect the phase differences between the four retrieved signals.}
    \label{fig:RT_QPSK_IQ}
\end{figure}

To showcase the excellent phase retrieval of Brillouin-based memory, we store and retrieve the 2-bit QPSK signals $\QPSKbit$. These are encoded as data pulses with the phases $\QPSK$. For our proof of principle experiment, we coherently store and retrieve a single data pulse at the \mbox{chosen} phase. However, as has already been demonstrated, \cite{Zhu2007,Kalosha2008,Merklein2017,Stiller2023} Brillouin-based memory is not limited to streams of single pulses. 

Another advantage of Brillouin-based memory is that it does not require extremely specialized equipment, but only standard telecom equipment and can be operated at room temperature. For this reason, we demonstrate phase retrieval of the four QPSK signals at room temperature with a storage time of $\tstore = 6$~ns. The BFS of the HNLF sample at room temperature is $\Omega/2\pi = 9.729$~GHz.

Unlike a previous work, \cite{Geilen2023} we do not trigger the measurements with the chosen $\phiD$ so the calibration pulse has a set phase, but allow it to rotate over the full phase space. We recover the phase of the
calibration pulse and use it to rotate every trace so that we are
able to obtain the mean for each $\phiD$. We also correct for errors from the double balanced homodyne detection using the Heydemann correction.\cite{Heydemann1981} Highlighted in \cref{fig:traces} are the mean $I(t)$ and $Q(t)$ traces for the 01 signal, showing the calibration pulse, data pulse, and readout. The mean reference trace --- obtained from traces taken with the control pulses off --- is given by dashed lines. This allows a phenomenon of Brillouin-based memory to be observed: directly after the depleted data pulse, an additional small peak can be seen that was not part of the initial data pulse. This has been termed the ``echo'' and comprehensively investigated by \citet{Dong2023} It is due to a pre-reading process as the control write pulse interacts with the already generated acoustic grating. 

For analysis in the IQ space, we calculate the area under the curve (AuC) of the data and readout pulses in the mean $I(t)$ and $Q(t)$ oscilloscope traces and normalize this to the AuC of the undepleted data pulse, which we find from the mean reference trace. Hence, the coordinates on the IQ diagrams in \cref{fig:RT_QPSK_IQ} are given by $\lb(\AuCI, \AuCQ\rb)$. The intrinsic global phase shift of $\pi$ of the readouts is shown in \cref{fig:RT_QPSK_IQ}(a), while the preservation of the relative phase shifts between the four QPSK signals is highlighted in \cref{fig:RT_QPSK_IQ}(b).

A common measure used in Brillouin-based memory are the storage and readout efficiencies, where the AuC is appropriate as it is proportional to the energy contained in the pulse. The storage efficiency is given by \mbox{$\lb(1 - \nicefrac{\text{AuC}_\text{\,depl. data}}{\text{AuC}_\text{\,init. data}}\rb)\times 100$} and readout efficiency is given by \mbox{$\lb(\nicefrac{\text{AuC}_\text{\,readout}}{\text{AuC}_\text{\,init. data}}\rb)\times100$}. For $\QPSKbit$, we achieve storage efficiencies of \{65\%, 64\%, 64\%, 66\%\} and readout efficiencies of \{41\%, 35\%, 36\%, 38\%\}, respectively. Theoretically, \cite{Zhu2007,Dong2015} close to 100\% storage and readout efficiency can be achieved. However, in practice, the finite acoustic lifetime makes such a high readout efficiency hard to reach.  

Further details of the data analysis is given in the supplementary materials, \cref{ssec:dataanalysis}.

%% file: cryoQPSK.tex
Augmenting our study at room temperature, we turn to measurements at cryogenic temperatures. At cryogenic temperatures, the lifetime of the acoustic wave is much longer than at room temperature.\cite{Geilen2023,LeFloch2003} We make use of this to validate that the QPSK signals can still be faithfully stored and retrieved at extended storage times, as shown in \cref{fig:4Kstorage}. There it is clearly seen that the extended storage times do not hinder the phase retrieval and that the readouts retain their relative phase shifts, as expected.\cite{Geilen2023} We see that at different storage times, the phase shift of the readout is slightly different. This could be due the measured BFS of the sample at 4.0~K ($\Omega/2\pi = 9.5987$~GHz) used to set $\omegaC$, being slightly off the exact BFS of the sample. In addition, the longer the storage time the more sensitive $\phiR$ is to the mismatch between $\omegaC$ and the BFS (see \cref{subsec:tunedcryoQPSK}).

With the longer acoustic lifetime also comes enhanced readout efficiency.  For $\QPSKbit$ at 6~ns, we achieve consistently higher readout efficiencies of \mbox{\{49\%, 48\%, 49\%, 49\%\}}, respectively. The storage efficiencies, \mbox{\{62\%, 62\%, 58\%, 58\%\}}, are relatively unchanged.  

For 30~ns and 70~ns we achieve storage efficiencies of \mbox{\{35\%, 33\%, 36\%, 33\%\}} and \mbox{\{35\%, 34\%, 34\%, 33\%\}}, respectively. The storage efficiencies for 6~ns are double due to the data pulse interacting with both write and read pulses i.e. being depleted twice. Nonetheless, the double interaction does not play a role in the readout process. For 30~ns and 70~ns we achieve readout efficiencies of \mbox{\{31\%, 30\%, 32\%, 31\%\}} and \mbox{\{15\%, 14\%, 11\%, 11\%\}}, where the decrease is expected (see \cref{subsec:lifetime}).

%% file: tunedcryoQPSK.tex
Penultimately, we experimentally demonstrate an interesting and valuable feature of Brillouin-based memory: complete control over $\phiR$ by manipulating the control pulses. When we are frequency matched, i.e. $\omegaC = \omegaD - \Omega$, the readout gains a $\pi$ phase shift with respect to $\phiD$. However, by detuning the control frequency, i.e. $\omegaC = \omegaD - \lb(\Omega \pm \DOmega\rb)$, we can arbitrarily tune the phase shift of the readout from $[0, 2\pi)$. To achieve a particular readout phase, the frequency detuning required depends on the chosen storage time. The relation between storage time and frequency detuning follows the relationship $\Delta\phiR \propto \DOmega\cdot\,\tstore$. Hence, for a storage time $\tstore^\pr = 5\cdot\tstore$, the required frequency detuning for the same phase in the readout will be $\DOmega^\pr = \frac{\DOmega}{5}$. This phase tuning can be performed any temperature and does not require large frequency detuning, reducing the requirements for its practical implementation. For example, at a storage time of 6~ns, a frequency detuning of only $\DOmega/2\pi$~=~75~MHz results in $\phiR~\approx~\phiD~+~2\pi$. 

 In the measurements of \Cref{subsec:cryoQPSK}, we additionally detuned $\omegaC$ by a range of frequencies --- enough for $\phiR$ to cover the full phase space. The range of frequency detuning was scaled based on the storage time, with a longer storage time corresponding to a reduced frequency sweep range.  The result (\cref{fig:detunedIQ}) was that we could run over the phase space in a controlled way, to have any arbitrary $\phiR$. For example, the outlined circles in \cref{fig:detunedIQ} highlight the control frequency detunings $\DOmega$ for which $\phiR = \phiD$. The outlined crosses highlight the frequency matched cases with $\DOmega~=~0$. For $\tstore$ = 6~ns, we see an anomalous decrease in the readout for positive frequency detunings. However, this is due to a modulator artifact rather than characteristic of Brillouin-based memory (see the supplementary materials, \cref{ssec:readout_anomaly}).

 This remarkable feature of Brillouin-based memory could prove to be a valuable tool in any future applications, for example real-time phase correction.

%% file: lifetimecryomemory.tex
\begin{figure}[t!]
    \centering
    \includegraphics{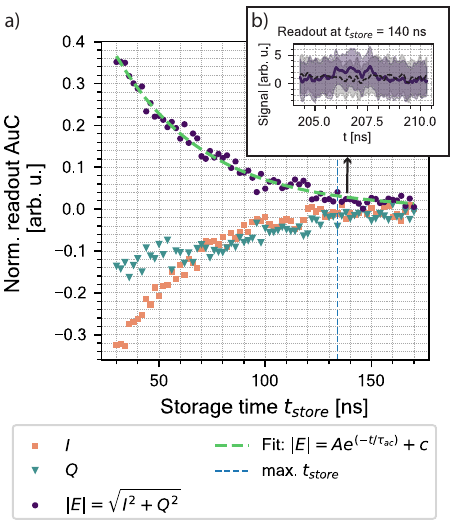}
    \caption{Decay of the readout AuC for storage times from 30 to 170~ns, measured for the 00 state with the control pulses frequency matched. (a) The total readout AuC, denoted $|E|$, is fitted with an exponential that decays proportional to the inverse of the acoustic lifetime, $\tau_{ac}$. Fit parameters: $A = 0.74 \pm 0.02$, $\tau_{ac} = 42 \pm 2 $ ns, $c = 0.000 \pm 0.006$. The values of $|E|$ are calculated from the $I$ and $Q$ AuCs. The vertical dashed line at 134~ns shows the position of the maximum storage time that is three standard deviations above the noise. (b) The inset shows the mean readout trace, in dark purple, for the furthest detectable storage time of 140~ns. The light purple shows one standard deviation above and below the mean readout trace. The mean reference trace is plotted as a dashed gray line and the light gray shows one standard deviation above and below the mean reference trace.}
    \label{fig:0_MHz_lifetime}
\end{figure}

The other side of the phase-tuning relation $\Delta\phiR \propto \DOmega\cdot\,\tstore$ can be seen when sweeping $\tstore$ while keeping $\DOmega$ fixed. Instead of picking a storage time and adjusting $\DOmega$, phase control could be achieved by picking $\DOmega$ and adjusting $\tstore$. We experimentally investigate the behavior of $\phiR$ for these circumstances by sweeping $\tstore$ from 30 to 170~ns for the 00 state for fixed $\DOmega$ values (\cref{fig:detunedlifetimeIQ}). When $\DOmega~\neq~0$, the readout decays to zero while $\phiR$ rotates either clockwise (positive detunings) or anti-clockwise (negative detunings). When $\DOmega~=~0$, the readout shows a simple decay to zero without a notable change in phase, as expected. The slight curve is for the same reason as \cref{fig:4Kstorage}, namely that we are not perfectly frequency matched with the sample resonance. 

Finally, a natural question arises when discussing storage techniques, namely, how long can signals be stored? Since the acoustic wave decays exponentially,\cite{Boyd2003} there is a limit to the storage time. To find this limit in our setup, we set our cryostat to the lowest possible temperature (3.9~K) for the storage time sweeps. 

Fitting the exponential decay of the total readout AuC with $\DOmega~=~0$ (\cref{fig:0_MHz_lifetime}(a)) reveals the acoustic lifetime to be \mbox{$\tau_{ac} = 42 \pm 2$ ns} ---  double the phonon lifetime since the phonon lifetime is measured using direct detection. The enhanced lifetime results in a detectable readout up to 140~ns, for which we can still retrieve the phase (\cref{fig:0_MHz_lifetime}(b)). Active refreshment schemes\cite{Stiller2020} could extend this number even further. Additionally, we consider a more conservative criterion for the maximum storage time based on the signal to noise ratio --- the largest $\tstore$ for which the readout is above three standard deviations of the mean noise\cite{Geilen2023} --- and find the maximum storage time to be 134~ns with this definition. 

Further details of the data analysis and additional plots are given in the supplementary materials, \cref{ssubsec:lifetimefits}.

%% file: conclusion.tex
\section{\MakeUppercase{Conclusion and discussion}}
\label{sec:conclusion}
We show that coherent optical information in the form of phase-shift keying techniques like QPSK can reliably be stored by Brillouin-based memory. Our scheme uses standard telecom equipment and is therefore straightforward to integrate into existing infrastructure. We have the ability to arbitrarily select the phase of the readout over the full phase space in a controlled manner, which is interesting for in-memory computing and signal cleaning. Notably, this phase control requires adjusting the control pulses alone and leaves the data unaltered. The scheme can be implemented at room temperature and cryogenic temperatures. At cryogenic temperatures, we can store the data pulses far longer, while still retaining the ability to coherently retrieve the information in the readout. At 3.9~K, we could coherently retrieve the data up to 140~ns after the original data was stored. 

At room temperature, the phase control of the readout can be achieved by offsetting the frequency of the control pulses from the BFS. At cryogenic temperatures, the access to extended storage times turns this into a two-part handle on the phase of the readout. Most importantly, the relative phases between QPSK signals are conserved. For a given storage time (control frequency), changing the control frequency (storage time) will add a fixed phase to all readouts.

%% file: drift.tex
\section{\MakeUppercase{Countermeasures for setup drift}}
\label{ssec:drift}
While most of the components of the setup had negligible drift, this was not true for the biases of the intensity quadrature modulator (IQ modulator) on the data arm. Hence, we needed to take measures to counteract this drift. 

A bias controller for the IQ modulator did not work in our case. While it kept the biases at a fixed point, the baseline was too high above zero. This limits the readout efficiency since the control read pulse interacts with the baseline noise as well as the acoustic wave. 

As an alternative, we limited the total run time. For the room temperature measurements in \cref{subsec:roomtempQPSK} in the main text, we measured a single storage time $\tstore$ for a single control frequency $\omegaC$. However, for the measurements at cryogenic temperatures in \cref{subsec:cryoQPSK,subsec:tunedcryoQPSK,subsec:lifetime} in the main text, we were additionally sweeping frequency and storage time. Hence these runs took much longer. For this reason, we took fewer total runs per data phase in the measurements at cryogenic temperatures than in the room temperature measurements.  For each data phase, $\phiD$, at room temperature (\cref{subsec:roomtempQPSK} in the main text), we took 300 measurements with 100 reference traces (i.e. traces with the control off, showing the undepleted data pulse). For each $\phiD,\ \tstore$ and $\omegaC$ at cryogenic temperatures (\cref{subsec:cryoQPSK,subsec:tunedcryoQPSK,subsec:lifetime} in the main text), we took 15 measurements with 15 reference traces.  

Lastly, to counter setup drift introducing systematic error, we randomized the order of the frequency and storage time sweeps in each of the 15 measurements for \cref{subsec:cryoQPSK,subsec:tunedcryoQPSK,subsec:lifetime}. 

%% file: dataanalysis.tex
\section{\MakeUppercase{Data analysis}}
\label{ssec:dataanalysis}

\input{BHDcali}
\input{AuCs}

\input{findingphases}

\input{lifetimecalcs}

%% file: BHDcali.tex
\subsection{Corrections of double balanced homodyne data}
\label{ssubsec:BHD}
For double balanced homodyne detection (DBHD), an optical hybrid (OH) and two balanced detectors (BDs) are needed. Ideally, the BDs would be identical in every way. In practice, the BDs will have slightly different gains. This means that the $I(t)$ oscilloscope trace will have artificially higher signal than $Q(t)$, or vice versa. The $I(t)$ and $Q(t)$ traces will, in practice, also be artificially offset from zero due to one input arm of the BD having higher power than the other input arm, i.e. imperfect balancing. Lastly, the OH may not produce a phase shift of exactly $\frac{\pi}{2}$. We correct for these three factors using the Heydemann correction.\cite{Heydemann1981} To find the correction factors, we take a large number of measurements of the undepleted data pulse for a given $\phiD$, enough so that the entire phase space is covered. 

We append all the single traces together to create one long trace with all the undepleted data pulses. Plotting this on a IQ diagram gives a filled ellipse. Then we apply an elliptical mask to select out the pulses and not the noise level of the appended trace. The mask selects everything that is above 80\% of the maximum pulse height i.e. the one axis selects everything in $I(t)$ above 80\% of the maximum of $I(t)$ and likewise for $Q(t)$ in the other axis. 

\begin{figure}[h]
    \centering
    \includegraphics{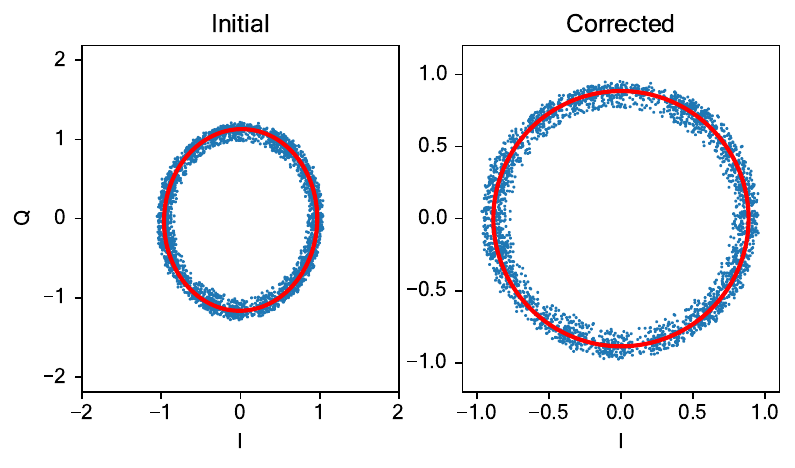}
    \caption{Calibration data before the Heydemann correction (``Initial'') and after (``Corrected''). The imperfect devices result in an ellipse in phase space, which are corrected to be circular.}
    \label{fig:HeydemannBHD}
\end{figure}

Using least squares regression, we fit an ellipse of the form 
\begin{equation}
    A x^2 + C xy + B y^2 + D x + E y = 1,
\end{equation}
where $x$ corresponds to $I(t)$ and $y$ corresponds to $Q(t)$. The fitting coefficients are then used to correct $I(t)$ and $Q(t)$ with the formulas given by \citet{Heydemann1981} to create a circular IQ plot. \cref{fig:HeydemannBHD} illustrates how the calibration data appears before and after applying the Heydemann correction. Before the corrections, the calibration data is elliptical in phase space whereas after the correction the calibration data is circular. 

We took 500 measurements for each run in \cref{subsec:roomtempQPSK} in the main text, and we took 250 measurements for each run in \cref{subsec:cryoQPSK,subsec:tunedcryoQPSK,subsec:lifetime} in the main text. We do this for each $\phiD$ since the imbalance between the detector arms drifts slightly over time. In addition, for each $\phiD$ we reset the IQ modulator biases and this creates slighlty different pulses. 

%% file: AuCs.tex
\subsection{Analysis using the area under the curve (AuC) of the pulses}
\label{ssubsec:AuC}

For all our analysis of the pulses in the IQ space, we use the area under the curve (AuC) of each pulse. To calculate the AuC, we select a set time window around the pulse in the $I(t)$ and $Q(t)$ traces. Then we sum all the points in that window to get $\AuCI$ and $\AuCQ$. The time window we use for each pulse is the same length and is equal to 6~ns, three times the data pulse width. This gives us a set of coordinates $\lb(\AuCI, \AuCQ\rb)$ in the IQ space. To find the total AuC, we can convert this to polar coordinates to get 
\begin{align}
    \text{AuC} = \sqrt{\AuCI^2 + \AuCQ^2}. \label{eq:AuC}
\end{align}
The phase can be determined from 
\begin{align}
    \phi = \arctan\lb({\frac{\AuCQ}{\AuCI}}\rb).
    \label{eq:angles}
\end{align}

%% file: findingphases.tex
\subsection{\MakeUppercase{Rotating the traces with respect to the calibration pulse}}
\label{ssubsec:findphases}

In order to account for the fact that, in fibers, the relative phase between the local oscillator (LO) and signal from the DBHD is not fixed but fluctuates in time, we have an additional pulse on the data arm: the calibration pulse. This pulse precedes the data pulse and acts as a phase reference for calibration. Using the arbitrary waveform generator (AWG) and IQ modulator, we choose the phase of the calibration pulse to be zero and the phase of the data pulse to be one of $\{0,\frac{\pi}{2},\pi,\frac{3\pi}{2}\}$. Hence, by calculating the angle of the calibration pulse ($\phi_\text{\,cali}$) using \cref{eq:angles}, we can use this angle to remove the global rotation in the calibration, data and readout pulses. In other words, we apply a rotation matrix, $\text{R}(-\phi_\text{\,cali})$ to the traces $I(t)$ and $Q(t)$:
 \begin{align}
     \begin{bmatrix}
        I^\pr(t) \\
        Q^\pr(t)
        \end{bmatrix} = 
        \text{R}(- \phi_\text{\,cali}) 
            \begin{bmatrix}
                I(t) \\
                Q(t)
            \end{bmatrix} = 
            \begin{bmatrix}
                    \cos(- \phi_\text{\,cali}) & -\sin( - \phi_\text{\,cali}) \\
                    \sin( - \phi_\text{\,cali}) & \cos( - \phi_\text{\,cali})
                \end{bmatrix} 
            \begin{bmatrix}
                    I(t) \\
                    Q(t)
            \end{bmatrix}
            \label{eq:rotmatrix}
 \end{align}
 so that the calibration pulse lies on the positive $I$ axis in the IQ plot.

\begin{figure}[t]
    \centering\includegraphics{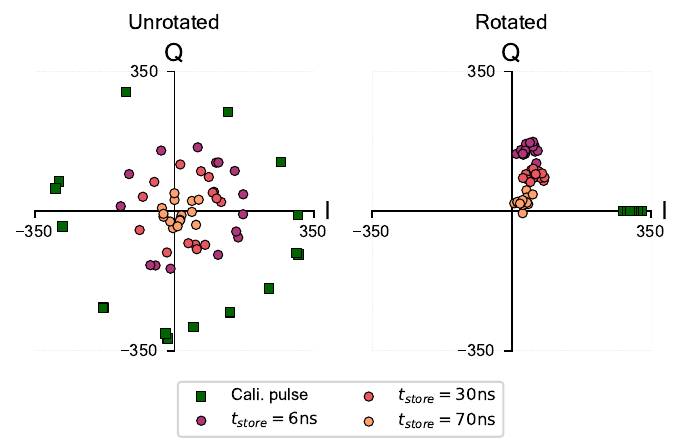}
    \caption{Before rotating the traces, the phases of the calibration and readout pulses range from $[0,2\pi)$ due to the fluctuating phase difference between the local oscillator and the signal. This creates a global rotation of the phases in the signal. However, since the calibration pulse is set to have a zero phase, we can remove the global rotation from each trace. The result is that the calibration pulses lie on the positive $I$ axis and the readout AuCs and phases can then be determined after obtaining the mean trace.}
    \label{fig:rotatinginIQ}
\end{figure}
 
For each trace, we find $\phi_\text{\,cali}$ and apply \cref{eq:rotmatrix}. Once we have rotated all the traces for a given data phase, storage time and control frequency, we can find the mean and standard deviation. This allows us to calculate the AuCs and phases from the mean trace, as described in \cref{ssubsec:AuC}. Before rotating, the mean would simply be zero since the relative phase between LO and signal will cover the entire phase space over time. The IQ plot of the calibration and readout AuCs for rotated and unrotated data is shown in \cref{fig:rotatinginIQ} where the readout phase is $\frac{\pi}{2}$ for the three storage times, $\tstore$, of 6, 30 and 70~ns.

All the AuCs of the readout and depleted data pulses are normalised with respect to the AuC of the undepleted data pulse. For the room temperature measurements in \cref{subsec:roomtempQPSK} in the main text, we obtain the AuC of the undepleted data pulse from 100 reference traces taken with the control arm off. For the measurements at cryogenic temperatures in \cref{subsec:cryoQPSK,subsec:tunedcryoQPSK,subsec:lifetime} in the main text, we obtain the AuC of the undepleted data pulse from 15 reference traces. 

We use Gaussian error propagation to calculate the accumulated error when calculating the $\AuCI$ and $\AuCQ$, when using \cref{eq:AuC} and when normalising the AuCs of the mean traces.

%% file: lifetimecalcs.tex
\subsection{Finding the maximal storage time}
\label{ssubsec:maxstore}
We define the maximum storage time as the largest $\tstore$ for which the readout is above three standard deviations of the mean noise. 

Firstly, we need to find the mean noise level and the corresponding standard deviation. This is done in two steps. For step one, we take a section at the beginning of the each trace, i.e. before the calibration pulse, where there is only noise and no possible other effects. With this noise section we calculate the mean and standard deviation of the noise for that trace. For step two, we find mean noise level of all the traces by taking the mean of the mean noise from each trace. We use Gaussian error propagation to calculate the corresponding standard deviation. 

\begin{figure}[H]
    \centering
    \includegraphics{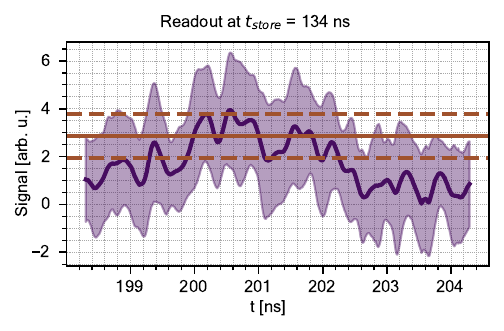}
    \caption{Mean readout trace, in dark purple, for a storage time of 134~ns --- the maximum storage time that met the criteria based on the mean noise. The light purple surrounding the mean readout trace shows one standard deviation above and below the mean readout trace. The solid brown line shows the obtained mean noise value and the dashed lines show the values that are three standard deviations above and below the mean noise value.}
    \label{fig:readout_134ns}
\end{figure}

After obtaining the mean noise level and the corresponding standard deviation, we compare the height of the readout in the mean trace to the value equal to three standard deviations above the mean noise level. We do this for all the storage times and find the largest $\tstore$ that satisfies our definition. At 3.9~K, we find this to be 134~ns (\cref{fig:readout_134ns}).

\subsection{Fitting the storage time sweep data of \cref{subsec:lifetime} in the main text}
\label{ssubsec:lifetimefits}

\begin{figure}[h]
    \centering
    \includegraphics{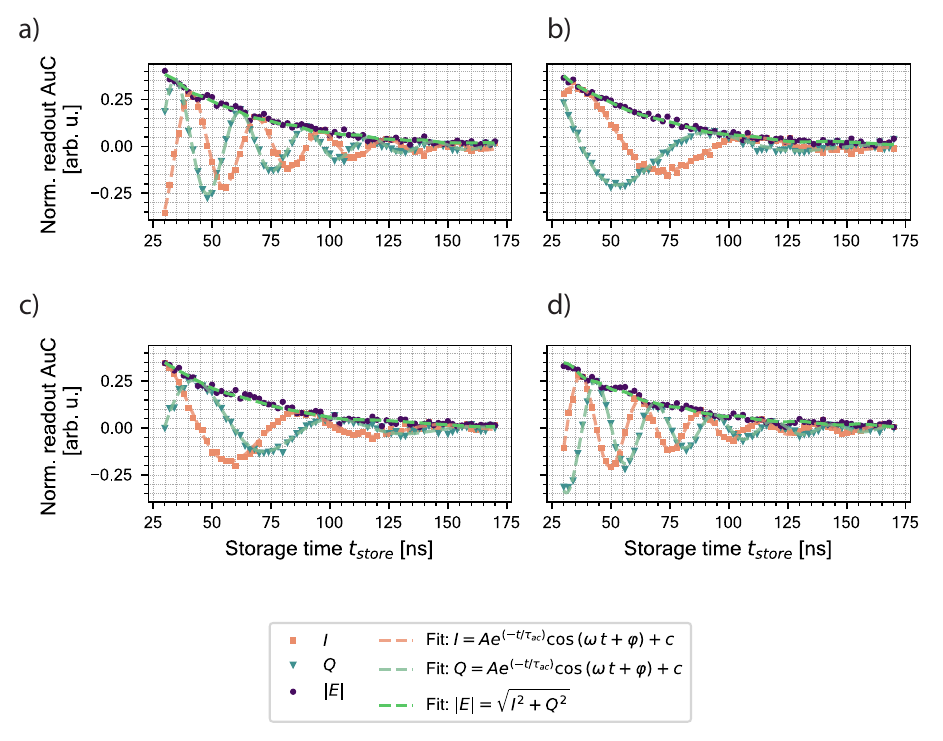}
    \caption{When the control pulses are frequency-detuned by $\Delta\Omega$, the readout does not decay as a simple exponential as in \cref{fig:0_MHz_lifetime} in the main text. Instead it has additional oscillations, which are prominent when looking at $\AuCI$ $\lb(\AuCQ\rb)$ as a function of storage time, $\tstore$. Above are measurements of the 00 state as $\tstore$ goes from 30 to 170~ns with the control pulses frequency detuned at a fixed value of (a) $\Delta\Omega$~=~-38~MHz (b) $\Delta\Omega$~=~-15~MHz (c) $\Delta\Omega$~=~-15~MHz and (d) $\Delta\Omega$~=~+38~MHz. In the plots,  $\AuCI$ $\lb(\AuCQ\rb)$ is denoted simply $I$ $\lb(Q\rb)$ and they are fitted as damped harmonic oscillators. \Cref{tab:fitparams} gives the fitting parameters. The fit function for the total AuC (denoted $|E|$) is derived from the fits of $I$ and $Q$, i.e. $|E|_\text{\,fit} = \sqrt{I_\text{\,fit}^{\ 2} + Q_\text{\,fit}^{\ 2}}$.}
    \label{fig:detunedlifetime}
\end{figure}

We use least squares regression to fit the data. For the case where we are frequency matched, we fit an exponential decay $A\, \exp\lb(\shortminus\,\nicefrac{t}{\tau_\text{ac}}\rb) + c$ to the total AuC of the readout and extract the acoustic lifetime $\tau_\text{ac}$. For the case where we are not frequency matched, we fit the oscillations in $\AuCI$ and $\AuCQ$ of the readout as damped harmonic oscillators $A\,\exp\lb(\shortminus\,\nicefrac{t}{\tau_\text{ac}}\rb)\, \cos\lb(\omega\, t + \varphi\rb) + c$. The fit for the total AuC of the readout shown in the plots of \cref{fig:detunedlifetime} is derived from the fits of $\AuCI$ and $\AuCQ$ i.e. AuC$_\text{\,fit} = \sqrt{\text{\,AuC}_\text{I,\,fit}^{\ 2} + \text{\,AuC}_\text{Q,\,fit}^{\ 2}}$. The fit parameters for the unmatched measurements are given in \cref{tab:fitparams}.

In \cref{tab:fitparams}, we see that the acoustic lifetime $\tau_\text{ac}$ is consistent with the value of 42~$\pm$~2~ns found for the frequency matched case in \cref{fig:0_MHz_lifetime} of the main text. We also see that the ratio of $\omega$ between the $\pm$~38~MHz and $\pm$~15~MHz is similar to the ratio $\frac{38}{15}$ of the frequency detuning. This matches that the frequency of the oscillations, $\omega$, is proportional to the detuning $\Delta\Omega$. Lastly and in line with expectations, we see that the $\cos$ functions in the $\AuCI$ and $\AuCQ$ fits are about $\frac{\pi}{2}$ apart, indicating a $\cos$ and $\sin$ pair.

\begin{table}[h]
\caption{Fit parameters for $\AuCI$ and $\AuCQ$ with fixed frequency detuning $\Delta\Omega$ MHz of the control pulses, while storage time was swept between 30 and 170~ns. In the table, $I$ $\lb(Q\rb)$ represents $\AuCI$ $\lb(\AuCQ\rb)$.}
\label{tab:fitparams}
\begin{ruledtabular}
\begin{tabular}{ccccccc}
    $\Delta\Omega$ (MHz) & & $A$ & $\tau_\text{ac}$ (ns) & $\omega$ (MHz) & $\varphi$ (rad) & $c$\\
    \hline \hline
     - 38 & $I$ & $0.76 \pm 0.3$ & $43 \pm 2 $& $227 \pm 1$& $-9.46 \pm 0.04$& $-0.004 \pm 0.002$\\
      & $Q$ & $-0.82 \pm 0.03  $&$41 \pm 1 $& $229 \pm 1$ & $-11.08\pm 0.04$ &$0.000 \pm 0.002$ \\
     - 15 & $I$ & $-0.77 \pm 0.03$ & $41 \pm 1$ & $-84.0 \pm 1.0$ & $6.36 \pm 0.05$ & $-0.003  \pm 0.002$ \\
      & $Q$ & $0.94  \pm 0.06$ & $39  \pm 2$ & $-82.0  \pm 1.0$ & $7.81  \pm 0.05$ & $0.001  \pm 0.002$ \\
     + 15 & $I$ & $-0.72 \pm 0.03$ & $42 \pm 1$ & $105.0 \pm 1.0$ & $-6.27  \pm 0.06$ & $-0.002  \pm 0.002$ \\
     & $Q$ & $-0.79  \pm 0.05$ & $36  \pm1$ & $106.0  \pm 1.0$ & $-7.84  \pm 0.05$ & $-0.005  \pm 0.002$ \\
     + 38 & $I$ & $-0.69 \pm 0.04$ & $42 \pm 2$ & $248.0 \pm 1.0$ & $-12.48  \pm 0.04$ & $-0.001  \pm 0.002$ \\
     & $Q$ & $0.72  \pm 0.03$ & $42 \pm 1$ & $-250.0  \pm 1.0$ & $11.02  \pm 0.05$ & $-0.005  \pm 0.002$ \\
\end{tabular}
\end{ruledtabular}
\end{table}

%% file: readoutanomalystudy.tex
\newpage
\section{Anomaly in frequency-detuned measurements at a storage time of 6~ns}
\label{ssec:readout_anomaly}

In \cref{subsec:tunedcryoQPSK} of the main text, the measurements of fequency detuned memory at $\tstore$ = 6~ns show an anomalous decrease in the readout for positive frequency detunings, compared to the consistent amplitude seen for the storage times of 30~ns and 70~ns. In this section, we show that this is due to a modulator artifact. We use a data phase $\phiD = 0$ for the example plots in the discussion. 

\begin{figure}[H]
    \centering
    \includegraphics{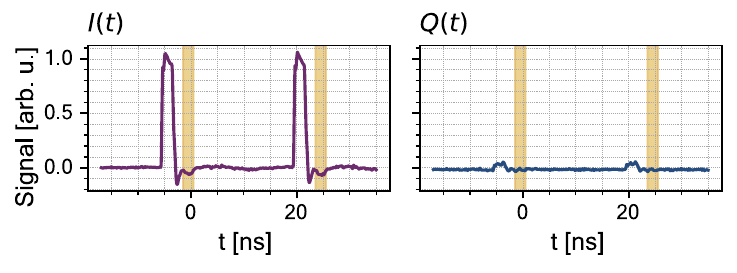}
    \caption{Example mean reference trace for $\phiD = 0$ for frequency detuned memory at 6~ns. Highlighted in yellow is the modulator artifact that can be seen directly after the calibration and undepleted data pulse in $I(t)$. Since this is for $\phiD = 0$, only a small portion of the pulses and artifacts are visible in $Q(t)$.}
    \label{fig:reference_traces}
\end{figure}

Directly after the pulses, there is an additional small bump created by the modulator. It is not due to any SBS effect since it can be seen in the calibration pulses and in the undepleted data in the reference traces. In \cref{fig:reference_traces}, the mean reference trace is plotted showing the calibration pulse and initial, undepleted, data pulse. The modulator artifact is highlighted in yellow.

\begin{figure}[h]
    \centering
    \includegraphics{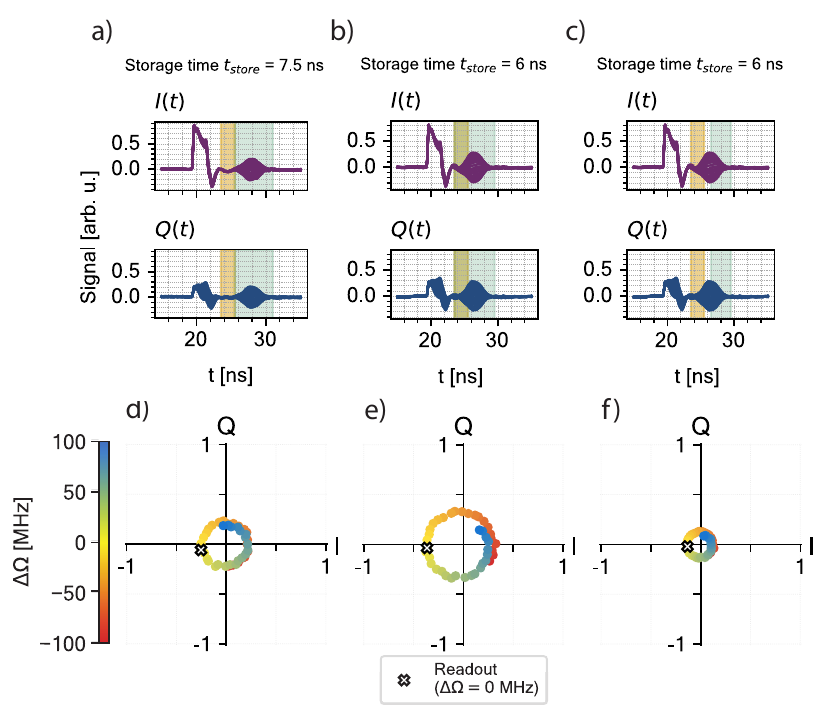}
    \caption{(a) Partial traces (cut to the data and readout pulse region) for a storage time $\tstore$~=~7.5~ns. The modulator artifacts (yellow), do not overlap with the readouts (green). (b) Partial traces for $\tstore$~=~6~ns, where the modulator artifacts overlap with the readouts. (c) The same partial traces as in Panel (b), but with only the half of the readouts selected that do not overlap with the modulator artifacts. (d) The IQ plot of the readout AuC data corresponding to Panel (a), where the readout AuCs have the same amplitude across the detuning ($\DOmega$) values. (e) The IQ plot corresponding to Panel (b), where the readout AuCs have differing amplitudes for the larger detunings. (f) The IQ plot corresponding to Panel (c), where only the latter half of the readout is selected. Now the readout AuCs have consistent amplitudes for the larger detunings.}    
    \label{fig:readout_anomaly_plots}
\end{figure}

 In \cref{fig:readout_anomaly_plots}(a), the partial mean rotated traces for all frequency detunings for $\tstore$ = 7.5~ns are plotted, cut to show the depleted data pulses and readouts. The depleted data pulses are followed by the modulator artifact, in yellow, and the readouts are highlighted in green. It can clearly be seen that for 7.5~ns, the readout region does not overlap with the modulator artifact whereas for 6~ns (\cref{fig:readout_anomaly_plots}(b)), it does. The IQ plot of the readouts at 7.5~ns across all the frequency detunings  (\cref{fig:readout_anomaly_plots}(d)) does not have the anomaly seen in \cref{fig:readout_anomaly_plots}(e). Furthermore, the anomaly for 6~ns disappears if we plot only the latter half readouts --- the selected region is illustrated in in the green highlighted area of \cref{fig:readout_anomaly_plots}(c) --- where there is no overlap with the modulator artifact (\cref{fig:readout_anomaly_plots}(f)).

We also ensured that this anomaly was not due to other effects. We made sure that the shifted control pulses were not getting cut off by the narrow bandpass filter at the extremes of the frequency sweep and that the control pulses had the same power for the entire frequency sweep. We checked to see if doubling the range of frequency detuning produced a decrease in readout for positive frequency detunings for 30~ns and 70~ns, but it did not. We repeated the measurements in the room temperature sample and the anomaly was still present. This, and a few other minor checks, allows us to conclude that the anomaly seen for the frequency detuned memory measurements at a storage time of 6~ns was caused by the modulator artifact. 
\vspace{0.4cm}